\documentclass[aip,floatfix,showpacs,preprintnumbers,amsmath,amssymb,nofootinbib,superscriptaddress]{revtex4-2}

\usepackage{color}
\usepackage{graphicx}
\usepackage{dcolumn}    % Align table columns on decimal point
\usepackage{multirow}
\usepackage{bm}         % bold math
\usepackage{sidecap}
\usepackage{hyperref}   % use for hypertext links, including those to external documents and URLs
\usepackage{xcolor}
\usepackage{amsmath}
\usepackage{color,colortbl}
\usepackage{booktabs} 
\usepackage{hhline}
\usepackage{ulem}
\usepackage{hyperref} 
\usepackage{url}

\definecolor{dark-red}{rgb}{0.,0.,0}
\definecolor{dark-blue}{rgb}{0.,0.,1}
\definecolor{medium-blue}{rgb}{0,0,1}
\definecolor{gray}{rgb}{0.85,0.85,0.85}

\hypersetup{
    colorlinks, linkcolor={dark-red},
    citecolor={dark-blue}, urlcolor={medium-blue}
}

\begin{document}

\title{Nuclear binding energy predictions using neural networks: Application of the multilayer perceptron}

\author{Esra Y\"{u}ksel}
\email{eyuksel@yildiz.edu.tr}
\affiliation{Department of Physics, Faculty of Science and Letters, Yildiz Technical University, Davutpasa Campus, 34220, Esenler, Istanbul, Turkey}

\author{Derya Soydaner}
\email{derya.soydaner@msgsu.edu.tr}
\affiliation {Department of Statistics, Mimar Sinan Fine Arts University, Bomonti 34380, Istanbul, Turkey}

\author{H\"{u}seyin Bahtiyar}
\email{huseyin.bahtiyar@msgsu.edu.tr}
\affiliation {Department of Physics, Mimar Sinan Fine Arts University, Bomonti 34380, Istanbul, Turkey}

\date{\today}% 

\begin{abstract}
In recent years, artificial neural networks and their applications for large data sets have became a crucial part of scientific research.
In this work, we implement the Multilayer Perceptron (MLP), which is a class of feedforward artificial neural network (ANN), to predict ground-state binding energies of atomic nuclei. Two different MLP architectures with three and four hidden layers are used to study their effects on the predictions. To train the MLP architectures, two different inputs are used along with the latest atomic mass table and changes in binding energy predictions are also analyzed in terms of the changes in the input channel. It is seen that using appropriate MLP architectures and putting more physical information in the input channels, MLP can make fast and reliable predictions for binding energies of atomic nuclei, which is also comparable to the microscopic energy density functionals. 
\end{abstract}

\maketitle

\section{INTRODUCTION} 

One of the major research areas in nuclear physics is the nuclear mass (binding energy) predictions, especially for nuclei far from the stability line with extreme proton-neutron ratio. As the most fundamental property of nuclei, accurate nuclear mass measurements and theoretical predictions have vital importance not only for nuclear physics \cite{RevModPhys.75.1021} but also for nuclear astrophysics \cite{PhysRevC.92.035807,Mumpower_2015,Schatz_2017}.  Alongside with other nuclear properties (i.e., charge radii, separation energies, decay properties, etc.), the nuclear masses could provide information about the nucleon-nucleon interaction, shell, and pairing properties of nuclei, and its precise determination is also crucial for our understanding of the formation of chemical elements heavier than iron in the universe \cite{PhysRevLett.116.121101}.

In the last decades, nuclear mass measurements have gained acceleration with the developments in experimental facilities. According to the latest atomic mass table AME2016 \cite{Wang_2017}, the ground-state masses of 3435 nuclei have been measured. However, measurement of the ground-state  properties for nuclei close to the drip lines still stands as a challenge. Also, there exist large deviations in theoretical model calculations for nuclei close to the drip lines. Therefore, further studies are needed to make reliable predictions in these regions. Up to now, microscopic-macroscopic (mic-mac) \cite{PhysRevC.82.044304,WS3,FRDM2012} and microscopic models \cite{PhysRevLett.102.152503,PhysRevLett.102.242501,STOITSOV2006243,nature,PhysRevC.93.054310,PhysRevC.89.054320} have been employed for the mapping of the nuclear landscape. Although the microscopic models are more complete in terms of the physics behind them, much better results are obtained using the mic-mac models in nuclear mass predictions since their constants are determined using the experimental ground-state masses of nuclei. While the root-mean-square (rms) deviation of nuclear masses is generally high (several MeV) using the microscopic models, the FRDM2012 \cite{FRDM2012} predicts an rms deviation 0.57 MeV with respect to the AME2003 atomic mass table \cite{AUDI2003337}, and WS3 model gives even lower rms deviation (0.336 MeV) for nuclear masses \cite{WS3}.

Considering the microscopic models, the first complete nuclear mass table was based on the well-known Skyrme Hartree–Fock (HF) method in the non-relativistic framework \cite{GORIELY2001311}, and the root-mean-square error was obtained as 0.738 MeV using the 1995 Audi–Wapstra compilation \cite{AUDI1995409}. With further improvements, the HFB-31 mass model also gave a model error of 0.561 MeV for the measured mass of 2353 nuclei \cite{PhysRevC.93.034337}. Using the Gogny HFB method, the rms deviation was obtained as 0.798 MeV with respect to the experimental predictions of the 2149 nuclei \cite{PhysRevLett.102.242501}. A systematic study was also conducted on 6969 nuclei to predict nuclear properties using the relativistic mean-field model, and rms deviation for nuclear masses was obtained as 2.1 MeV \cite{10.1143/PTP.113.785}. According to the latest microscopic mass model based on the relativistic continuum Hartree-Bogoliubov (RCHB) theory \cite{XIA20181}, the root-mean-square deviation of the binding energies with respect to the experimental data was obtained at around several MeV. Although considerable progress has been achieved with the mic-mac and microscopic models, the rms deviations are still high and one needs more accurate results, especially for astrophysical applications. Therefore, different approximations and models are required to understand the discrepancies in the results as well as to make more precise predictions.

In recent years, there has been an increasing amount of interest in artificial neural networks \cite{Akkoyun_2013,BAYRAM2014172,Goodfellow-et-al-2016,PhysRevC.96.044308,NIU201848,PhysRevC.98.034318,PhysRevLett.122.062502}, which is known as a nonparametric estimator in Machine Learning (ML). Applications of neural networks cover many areas in science as well as the different branches of physics. Considering the variety and richness of the available experimental data, nuclear physics is also a good candidate to study using neural networks. Long ago, several studies were performed to predict nuclear properties using various techniques \cite{GERNOTH19931,GAZULA19921,ATHANASSOPOULOS2004222}. Neural networks were used to predict nuclear mass excess and neutron separation energies \cite{GAZULA19921}, and it is shown that neural networks can be used as a new tool to predict the properties of atomic nuclei. Later on, nuclear mass defect predictions were made using the neural networks, showing that the neural networks can be considered as powerful tools to explore nuclear properties alongside the theoretical models \cite{ATHANASSOPOULOS2004222}. The ground-state energies \cite{BAYRAM2014172} and charge radii \cite{Akkoyun_2013} of nuclei were also investigated using artificial neural networks, and the usefulness of the method in the predictions was shown. Recently, various machine learning algorithms were used with the latest AME2016 data set to estimate the binding energies of atomic nuclei \cite{machine}. Besides, it was shown that the deep neural networks can predict the ground-state and excited energies as accurate as the nuclear energy density functional with less computational cost \cite{PhysRevLett.124.162502}. In recent years, neural network approaches were also used to train the mass residues of the theoretical models to improve the predictive power of the models and achieved considerable success \cite {PhysRevC.96.044308,NIU201848,PhysRevC.98.034318,PhysRevLett.122.062502}. As far as we are concerned, there is no work related to the application of the multilayer perceptron (MLP), which is a class of feedforward artificial neural network (ANN), to nuclear physics data. Therefore, it would be interesting to investigate the success of this model in the predictions of nuclear properties.

In this work, we implement the multilayer perceptron to predict the total binding energies (BE) of atomic nuclei. In our work, we first use the experimental data \cite{Wang_2017} along with the proton (Z) and mass (A) numbers of the selected nuclei as inputs. Then, we study the effects of the increasing number of the hidden layers and inputs in the predictive power of the neural network. Finally, we compare our results with the other microscopic and microscopic-macroscopic models to evaluate the success of MLP in the binding energy predictions compared to the other models.

\section{Multilayer Perceptron} 
\label{sec:1}

In this study, we aim to create a model that makes ground-state binding energy predictions for atomic nuclei by using input data. Inputs are the nuclear properties that can affect the binding energies of nuclei. Our model takes these properties as input and predicts the binding energies as the output. Such problems, where the output is a numerical value, are known as the \textit{regression} problems. Regression is a \textit{supervised learning} problem where there is an input, \textit{X}, an output, \textit{Y}, and the task is to learn the mapping from the input to the output \cite{Alpaydin2014}. To this end, in machine learning, we assume a model as shown below:

\begin{equation}
y = f(x|\theta)
\end{equation} 
where $f(.)$ is the model and $\theta$ are its parameters. In our case, \textit{$y$} corresponds to the prediction for binding energy, and $f(.)$ is the regression function. In the context of machine learning, the parameters, $\theta$, are optimized by minimizing a loss function. Thus, the predictions are obtained as close as possible to the correct values given in the input data. 

We choose multilayer perceptron (MLP) as the model $f(.)$. MLP is a neural network architecture that is mostly preferred to solve such regression problems. In the training stage of an MLP, the backpropagation algorithm \cite{Rumelhart1986} is used for computing the gradient. On the other side, another algorithm is used to perform learning using this gradient \cite{Goodfellow-et-al-2016}. The second algorithm is used for optimization, which is usually called as the \textit{optimizer}. In recent years, a new type of algorithm, which is called the adaptive gradient method, is preferred as the optimizer. In this study, we use \textit{Adam} algorithm \cite{Kingma2014} to train the MLP. 

Basically, an MLP is a feedforward neural network with one or more than one hidden layer between input and output layers. In the case of one hidden layer, first, input \textit{$x$} is fed to the input layer. By using an activation function, the \textit{activation} propagates in the forward direction, and the values of the hidden units \textit{z} are computed. Each hidden unit usually applies a nonlinear activation function to its weighted sum. After performing the forward pass, an error is computed by using a loss function. By using this error, the weights are updated in the backward pass \cite{Rumelhart1986}. 

However, an MLP with one hidden layer has limited capacity, and using an MLP with multiple hidden layers can learn more complicated functions of the input. That is the idea behind \textit{deep neural networks} where each hidden layer combines the values in its preceding layer and learns more complicated functions \cite{Alpaydin2014}. It is possible to have multiple hidden layers each with its own weights and applying the activation function to its weighted sum. It should be noted that different activation functions can be used in multilayer perceptrons, e.g., ReLU, tanh, sigmoid, etc. In this work, we implement both tanh and ReLU, which are two commonly used activation functions in nuclear mass predictions \cite{idini,machine,Zhang_2017,NIU201848}, and find that the ReLU function gives better predictions on the test data. Therefore, we choose the ReLU function as the activation function of the hidden layers in this work, which is also mostly preferred for the hidden layers of deep neural networks \cite{Glorot2011}:
\begin{equation}
\phi(x) = max(0,x)
\end{equation}

\begin{figure}[htbp]
	\centering
	\includegraphics[width=0.8\linewidth,clip=true]{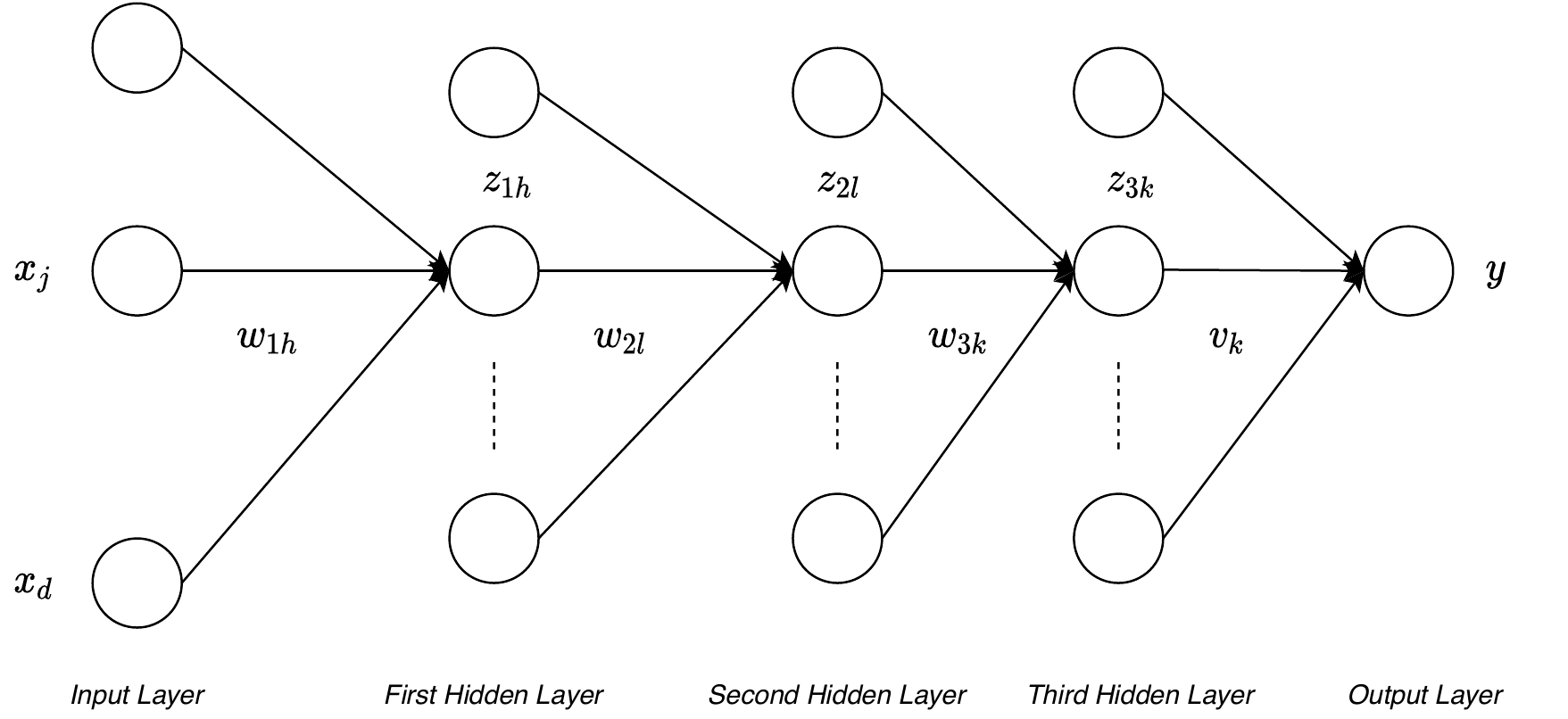} 
	\caption{The structure of a multilayer perceptron with three hidden layers.}
	\label{mlp}
\end{figure}

An MLP with three hidden layers is demonstrated in Fig. \ref{mlp} where \textit{$\textbf{w}_{1h}$}, \textit{$ \textbf{w}_{2l}$} and \textit{$ \textbf{w}_{3k}$} are the weights belonging to the first, second and third hidden layers, respectively. The units on the first, second and third hidden layers are represented as \textit{$z_{1h}$}, \textit{$z_{2l}$} and \textit{$z_{3k}$}, and \textbf{v} are the output layer weights. Such an architecture is required four stages to compute the output. Firstly, input \textit{x} is fed to the input layer, the weighted sum is computed, and the activation propagates in the forward direction. When the ReLU function is chosen as the activation function, \textit{$z_{1h}$} is computed as shown below:

\begin{eqnarray}
z_{1h} &=& ReLU( \textbf{w}_{1h}^T \textbf{x}) 
\nonumber\\
&=& ReLU\Bigg(\sum_{j=1}^d w_{1hj}x_j + w_{1h0} \Bigg) ,  h=1,...,H_1
\end{eqnarray} 

The computations for the second hidden layer are practiced similarly. At this stage, the second hidden layer activations are computed by taking the first hidden layer activations as their inputs. Then, the third hidden layer activations are computed by taking the second hidden layer activations as their inputs. In a regression problem, there is no nonlinearity in the output layer. Therefore, the output \textit{$y$} is computed by taking the \textit{$z_3$} as input \cite{Alpaydin2014}. Thus, the forward propagation is completed: 

\begin{eqnarray}
z_{2l} &=& ReLU( \textbf{w}_{2l}^T \textbf{z}_1) 
\nonumber\\
&=& ReLU\Bigg(\sum_{h=0}^{H_1} w_{2lh}z_{1h} + w_{2l0} \Bigg) ,  l=1,...,H_2
\end{eqnarray}

\begin{eqnarray}
z_{3k} &=& ReLU( \textbf{w}_{3k}^T \textbf{z}_2) 
\nonumber\\
&=& ReLU\Bigg(\sum_{l=0}^{H_2} w_{3kl}z_{2l} + w_{3k0} \Bigg) ,  k=1,...,H_3
\end{eqnarray}

\begin{equation}
y = \textbf{v}^T \textbf{z}_3 = \sum_{k=1}^{H_3} v_{k}z_{3k} + v_{0}
\end{equation} 

When the MLP goes deeper, one more step is added to these computations for each one of additional hidden layer. In this study, we implement MLP architectures of two different depths. Whereas the first one includes three hidden layers, the other one includes four. Thus, we observe the effect of depth on the binding energy predictions. As we predict the binding energy, i.e. one single numerical value, only one unit exists in the output layer. In order to determine the MLP architecture, we gradually increase the number of hidden units according to the prediction performance on three data sets. Then, we choose our final architectures. The MLP with three hidden layers includes 32,16 and 8 hidden units, respectively. It includes 769 (833) parameters for the MLP model with two (four) inputs, which is much smaller than the number of training data. On the other side, the MLP with four hidden layers includes 32,32,16 and 8 hidden units. It includes 1825 (1889) parameters for the MLP model with two (four) inputs. In addition to the smaller number of parameters, our architectures does not overfit because of two main reasons: Firstly, the central challenge in machine learning is that we must perform well on new, previously unseen inputs – not just those on which our model is trained \cite{Goodfellow-et-al-2016}. As it is seen in our results below, our neural network can make good predictions on test data. Secondly, overfitting occurs when the gap between the training loss and test loss is too large \cite{Goodfellow-et-al-2016}. However, in our calculations, the gap between the training loss and test loss is too small. For instance, it is found as 0.0023 (0.0029) for the training (test) data of MLP architecture with four hidden layers using four inputs. Therefore, it is seen that our neural network does not overfit, and it generalizes well on test data of each dataset, and the losses of training and test data are so close to  each other.

Another important step of creating these architectures is to initialize the layer weights. We initialize them with the Glorot normal initializer, also known as Xavier normal initializer \cite{Glorot2010}. Besides, the input data is randomly divided into two subsets as 70.0\% for training and 30.0\% for testing. We prefer \textit{mean absolute error} as the loss function on the training set \textit{X}:

\begin{equation}
E(\textbf{W},\textbf{v}|X) = \frac{\sum_{t=1}^n \left | r^t - f(x^t)\right |}{n} 
\end{equation}
where $r^t$ are the desired values and $f(x^t)$ are predictions for the binding energy. 

We train our MLP architectures 800 epochs by using Adam optimization algorithm to minimize mean absolute error. The name \textit{Adam} is derived from adaptive moment estimation. It is an adaptive gradient method that individually adapts the learning rates of model parameters \cite{Kingma2014}. During training, this algorithm computes the estimates of first and second moments of the gradients, and uses decay constants to update them. Therefore, Adam algorithm requires hyperparameters called \textit{decay constants} in addition to the \textit{learning rate}. In this study, the initial learning rate is 0.001, and decay constants are 0.9 and 0.999, respectively.

\section{Results}
\label{sec:2}

\begin{figure}[!ht]
	\begin{center}
	\includegraphics[width=0.8\linewidth,clip=true]{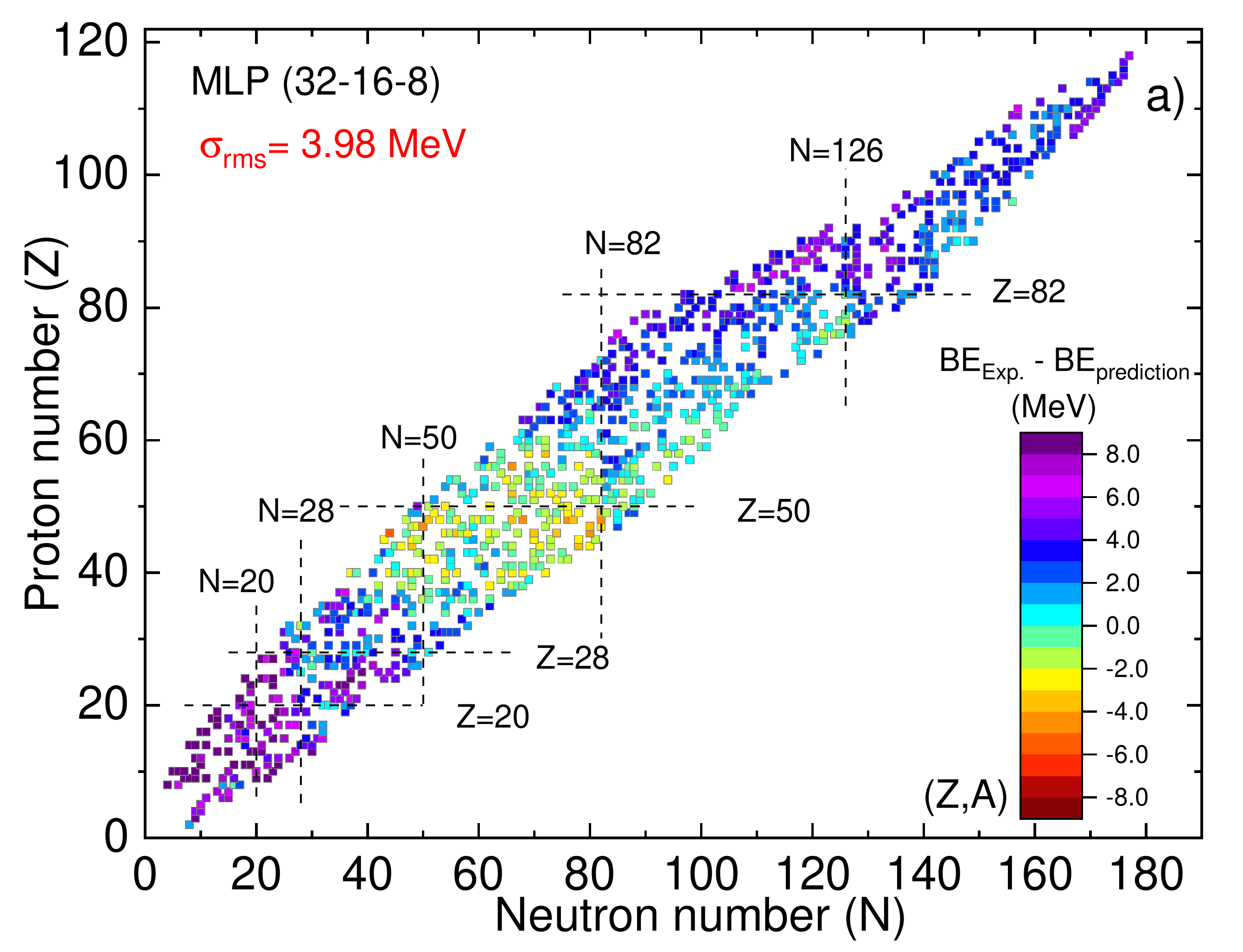} 
	\includegraphics[width=0.8\linewidth,clip=true]{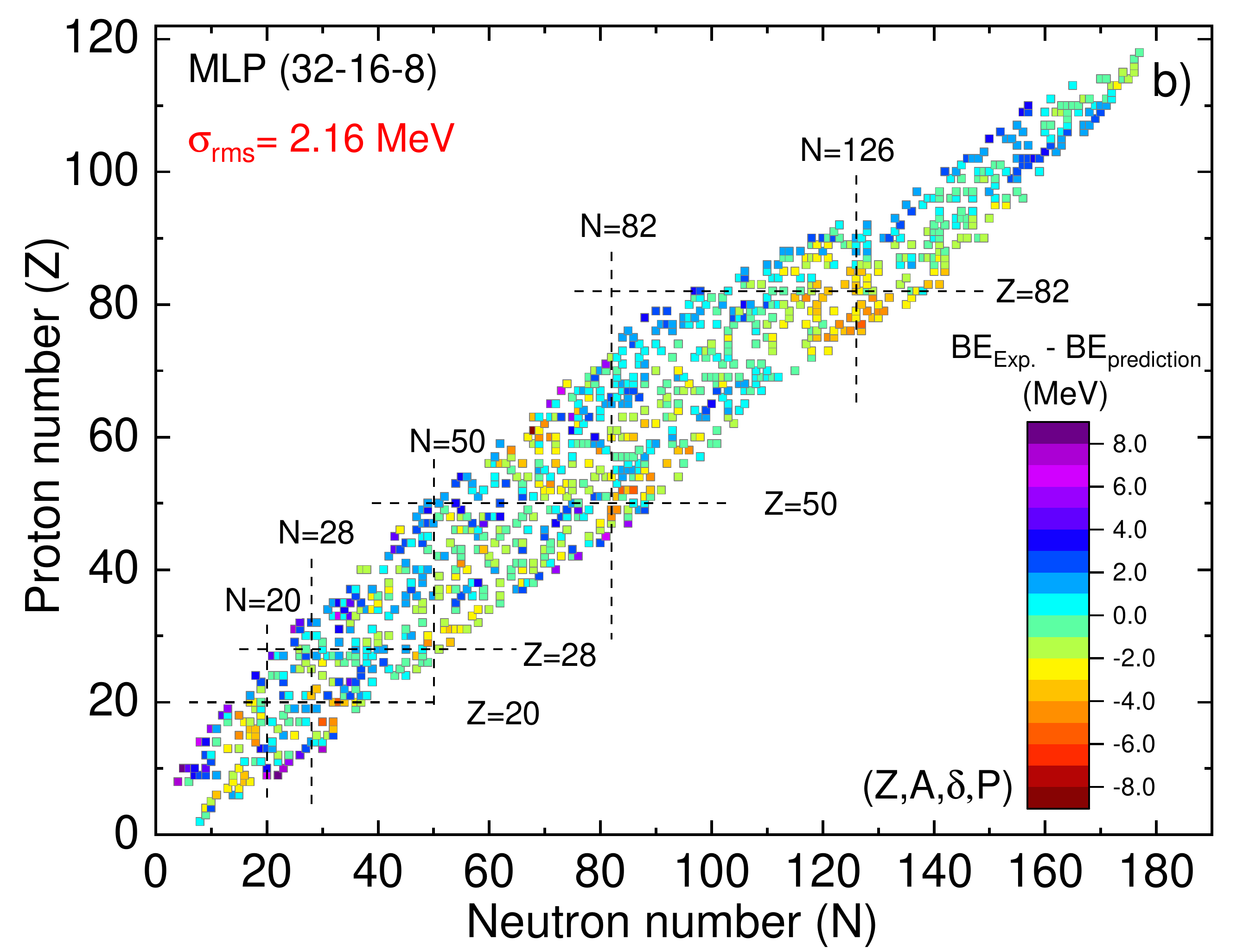} 
	\end{center}
	\caption{Comparison of the experimental and predicted binding energy differences for the testing set of the three layers MLP architecture using (Z, A) (upper figure) and (Z, A, $\delta$, P) (lower figure) as inputs.} 
	\label{mass0}
	\end{figure}
	\begin{figure}[!ht]
	\begin{center}
	\includegraphics[width=0.8\linewidth,clip=true]{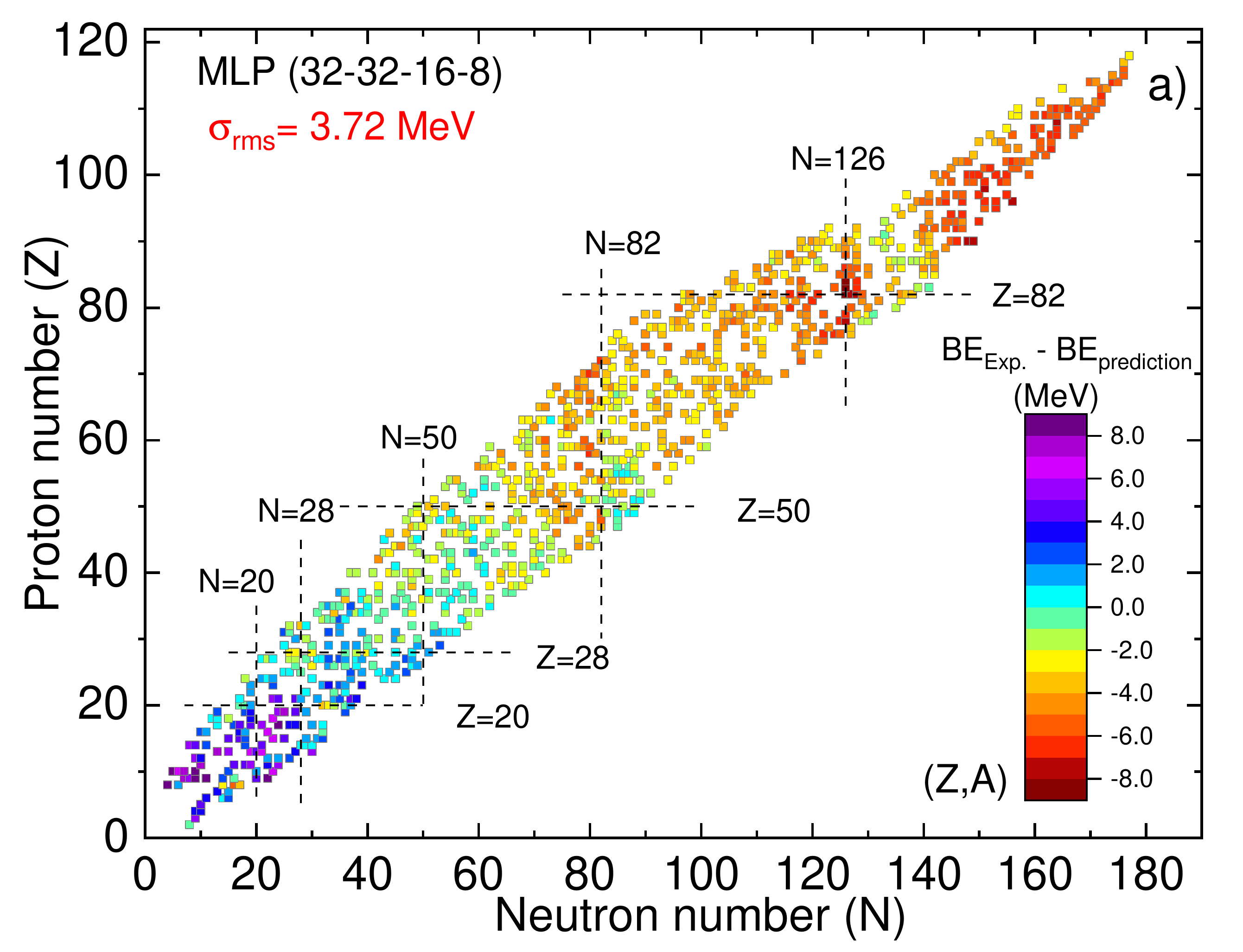} 
	\includegraphics[width=0.8\linewidth,clip=true]{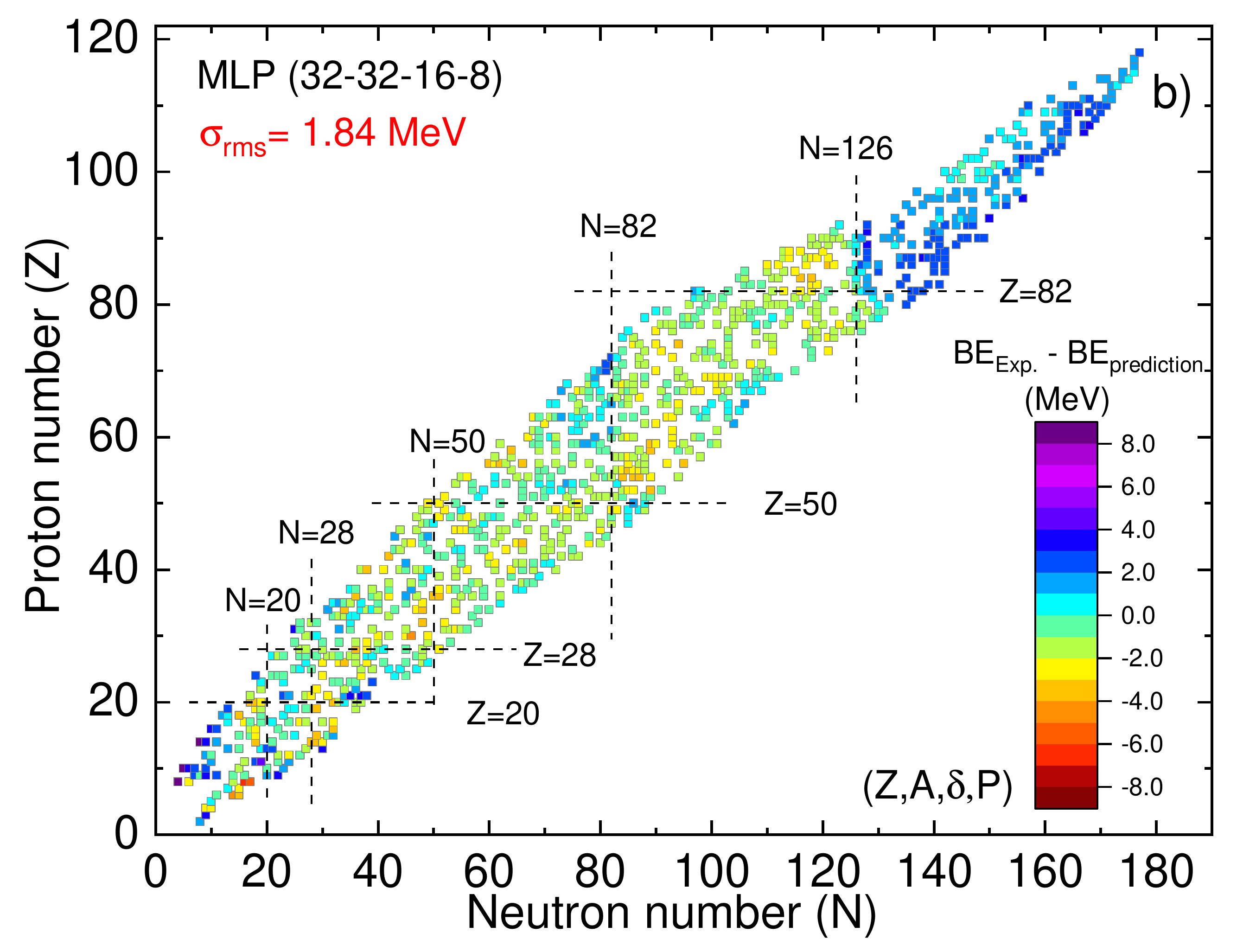} 
	\end{center}
	\caption{The same as in Fig. \ref{mass0}, but for the four layers MLP architecture.} 
	\label{mass1}
	\end{figure}

In this work, the multilayer perceptron is used to predict ground-state binding energies of atomic nuclei. Two different MLP architectures and inputs are used to test the effect of the number of the hidden layers and inputs on the results. In the input channels, we first use proton (Z) and mass (A) numbers of nuclei as inputs along with the experimental data from the latest AME2016 mass table \cite{Wang_2017} to predict binding energies. Therefore, this part of the present work does not include any physical identity or theory, except the proton and mass numbers. In the second part, we also include additional physical inputs to improve the predictive power of our models. These inputs carry information about the shell structure of nuclei, which in turn related to nuclear binding energies \cite{Casten_1996,KIRSON200829}. One of them is the pairing term $\delta(Z, N)$, which is defined as

\begin{equation}
\delta(Z, N) = \left[(-1)^{Z}+(-1)^{N}\right]/2.
\end{equation}

The pairing term becomes $+1$ ($-1$) for even-even (odd-odd) nuclei, and 0 for other nuclei. A positive value for the pairing term indicates that the nucleus is more bound while the opposite behavior is valid for a negative value \cite{KIRSON200829}. Another input is the promiscuity factor (P) of nuclei \cite{Casten_1996} and it is given by 

\begin{equation}
P = \nu_{p}\nu_{n}/( \nu_{p}+\nu_{n}),
\end{equation}

where $\nu_{p(n)}$ is the difference between the actual proton (neutron) number and the nearest magic number. The promiscuity factor is defined as a measure of the valance proton-neutron ($p-n$) interactions \cite{Casten_1996}. In this work, the proton and neutron magic numbers are taken as Z=8, 20, 28, 50, 82, 126 and N=8, 20, 28, 50, 82, 126, 184, respectively. The latest AME2016 mass table provides data for 3413 nuclei for A$\geq$8. However, only 2479 of them are obtained experimentally, and the others are calculated using the trend from the mass surface (TMS) in the neighborhood. Although the properties of some nuclei are not obtained directly from experiments, they are expected to provide reasonable data and trends for unknown nuclei. As it is well-known, having a large amount of data is crucial for training an artificial neural network. Using a large collection of data, performance of an algorithm can be improved significantly. In our model, we make predictions by using only proton and mass numbers of nuclei alongside some shell effects in the input. Since we do not provide much information to the model in the input channel, we need a large amount of data to make reasonable predictions. Therefore, non-experimental values from the AME2016 mass table are also used in the calculations to increase the performance of the model. While training the model, light nuclei with N$<$8 and A$<$10 is not taken into account and we randomly divide our data set (3388 nuclei in total) into training (70.0\%) and testing sets (30.0\%), as usual.

In Figs. \ref{mass0} and  \ref{mass1}, we display the binding energy (BE) differences between the experimental data and the results from the MLP model using two different architectures and inputs. The first architecture has three hidden layers (32-16-8), and we only use the proton and mass numbers of nuclei (Z, A) in the input channel (see Fig. \ref{mass0}(a)). Then, we also add pairing and promiscuity factors (Z, A, $\delta$, P) to see their effects on the results (see Fig. \ref{mass0}(b)). 

The predictions of the MLP with the three hidden layers are given in Fig. \ref{mass0} for 1017 nuclei in the testing set.
Although the root-mean-square deviation ($\sigma_{rms}$) with respect to the experimental data is high and obtained as $\sigma_{rms}=3.98$ MeV, the MLP can make reasonable predictions for the nuclear binding energies using only (Z, A) as inputs, and the results are comparable with the predictions of the nuclear energy density functionals. First thing to notice is the large deviation of the model results for light and heavy nuclei, which can be related to the limited number of experimental data in these regions.
Adding more physical information to the input, we find that the predictive power of the MLP increases considerably as can be seen from Fig.\ref{mass0}(b). By adding pairing and promiscuity factors to the input channel, the root-mean-square deviation is improved by about 45.73\% and obtained as 2.16 MeV for the testing set nuclei. It is also clear that the predictive power of the MLP increases considerably for light and heavy nuclei. To see the performance of the MLP in different regions of the nuclear landscape, we divide the testing set into three parts as light nuclei (Z$<$20, 93 nuclei), medium-heavy nuclei (20$\leq$Z$\leq$82, 696 nuclei), and super-heavy nuclei (Z$\geq$82, 228 nuclei). Using (Z, A) as inputs, the $\sigma_{rms}$ values are obtained as 8.60, 2.93, and 3.80 MeV for light, medium-heavy and super-heavy nuclei, respectively. Increasing the number of the inputs in the MLP architecture, namely using (Z, A, $\delta$, P) in the input channel, we obtain important improvements in the $\sigma_{rms}$ values (see Table \ref{table00}). For instance, the $\sigma_{rms}$ value for light nuclei is decreased and found as 3.39 MeV, which corresponds to 60.58\% improvement in the predictions. Besides, the $\sigma_{rms}$ value for medium-heavy and super-heavy nuclei is decreased and obtained as 2.10 and 1.61 MeV, respectively. Using MLP with four inputs, the model predictions are improved by about 28.32\% and 57.63\% for medium-heavy and super-heavy nuclei, respectively.
\begin{table*}
\caption{The root-mean square deviations ($\sigma_{rms}$) in units of MeV for different MLP architectures and inputs. The results of the best MLP architectures are shown in bold. Since the number of the parameters are higher than the number of the training data set, the results of the two hidden layers (64-64) MLP architecture are not presented.} 
\begin{center}
%\vspace{6mm}
\tabcolsep=0.5em \renewcommand{\arraystretch}{1.0}%
\begin{tabular}
[c]{ccccccccc}\hline\hline 
\\ [-1ex]
 $$&MLP& Input & Z$<$20 & $20\leq$ Z$\leq 82$ & Z$>$82& Testing set  \\
 $$&& &93 nuclei & 696 nuclei & 228 nuclei & 1017 nuclei  \\
\hline
 & (32-32) & (A,Z) &7.80 &2.80 &2.11& 3.46&   \\
 & (64-32) & (A,Z) &2.93 &2.35 &4.50&3.01&   \\
 & (32-32) & (A,Z,$\delta$,P) &4.13 &1.95 &4.76& 3.04&   \\
 & (64-32) & (A,Z,$\delta$,P) &5.00&1.88&3.07&2.61&   \\
\hline
 & (32-16-8) & (A,Z) &8.60 &2.93 &3.80& 3.98&   \\
 & (64-8-4) & (A,Z) &6.42 &4.07 &4.83& 4.51&   \\
 & (64-16-8) & (A,Z) &6.22 &2.11 &2.05& 2.75&   \\
 & \textbf{(32-16-8)} & \textbf{(A,Z,$\delta$,P) }&\textbf{3.39} &\textbf{2.10} &\textbf{1.61}& \textbf{2.16}&   \\
  & (64-8-4) & (A,Z,$\delta$,P) &6.75 &2.93 &4.76& 3.90&   \\
 & (64-16-8) & (A,Z,$\delta$,P) &4.60 &1.89 &2.47& 2.40&   \\
\hline 
 & (32-16-8-4) & (A,Z) &10.37 &3.01&2.60& 4.20&   \\
 & (32-16-16-8) & (A,Z) &1.98 &2.73 &2.37& 2.60&   \\
 & (32-32-16-8) & (A,Z) &4.89 &3.06 &4.82& 3.72&   \\
 & (32-16-8-4) & (A,Z,$\delta$,P) &5.03 &3.22 &3.24& 3.43&   \\
 & (32-16-16-8) & (A,Z,$\delta$,P) &9.11 &2.83 &2.27& 3.78&   \\
& \textbf{(32-32-16-8)} & \textbf{(A,Z,$\delta$,P)} &\textbf{3.03} &\textbf{1.58} &\textbf{1.94}& \textbf{1.84}&   \\
\hline 
 & (32-16-8-8-8) & (A,Z) &4.20 &2.56&5.25& 3.50&   \\
 & (32-16-16-8-4) & (A,Z) &4.21 &2.83 &4.88&3.52&   \\
 & (64-32-16-16-8) & (A,Z) &2.87 &2.46 &5.55&3.44&   \\
 & (32-16-8-8-8) & (A,Z,$\delta$,P) &5.56 &2.17 &1.35&2.54&   \\
 & (32-16-16-8-4) & (A,Z,$\delta$,P) &10.83 &2.98 &1.78& 4.18&   \\
 & (64-32-16-16-8) & (A,Z,$\delta$,P) &3.83 &7.47 &2.47& 4.92&   \\
\hline\hline
\end{tabular}\\ [-1ex]
\end{center}
\label{table00}
\end{table*}

\begin{figure*}[!ht]
	\begin{center}
		\includegraphics[width=0.85\linewidth,clip=true]{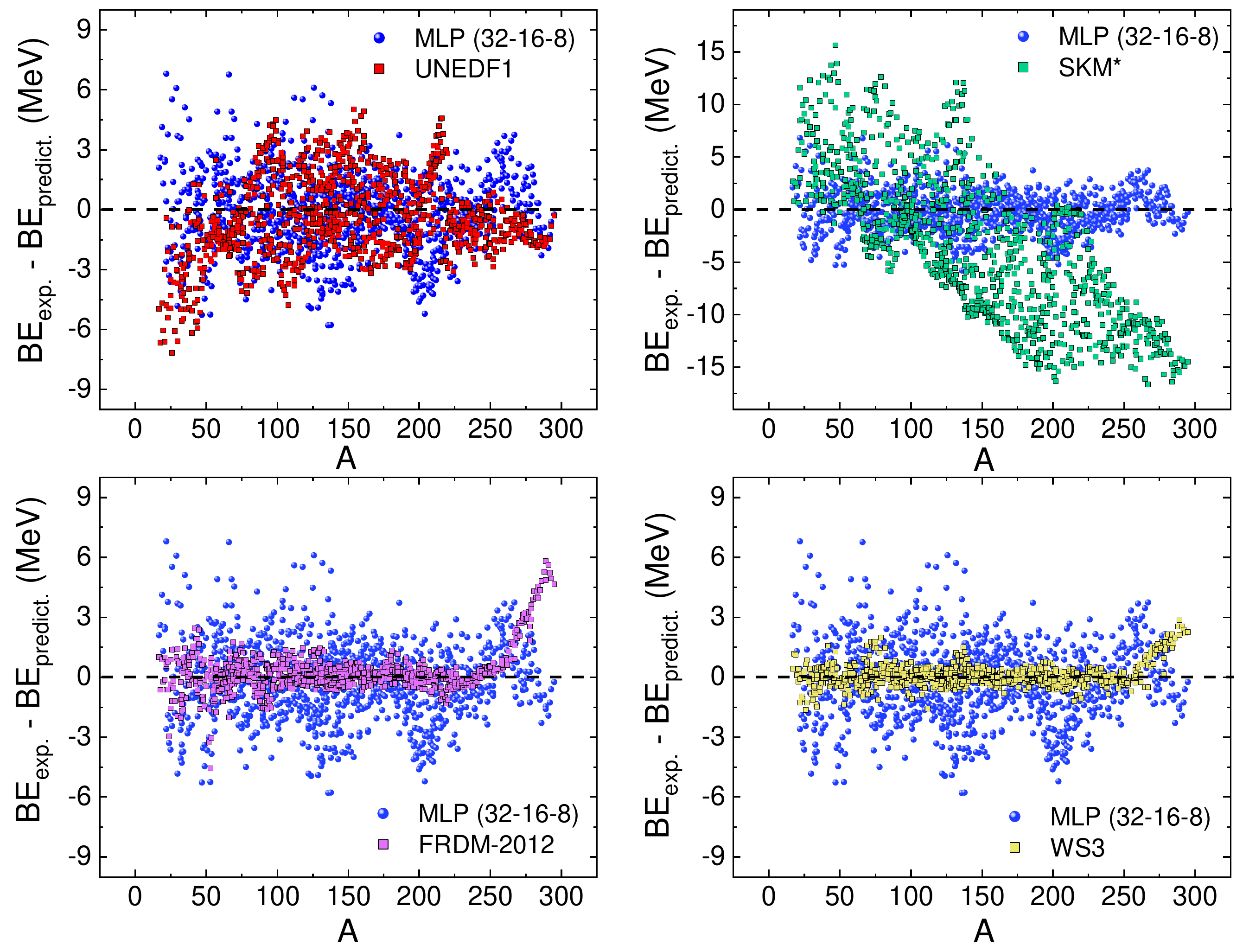}
	\end{center}
	\caption{Comparison of the experimental and predicted binding energy differences for 956 nuclei between the MLP with three hidden layers and (a) UNEDF1 \cite{UENDF1,masstables,nature}, SKM* \cite{SKM*,masstables,nature}, (c) FRDM-2012 \cite{FRDM2012} and (d) WS3 \cite{WS3} models. The black dashed lines are given to guide the eye.} 
	\label{mass}
\end{figure*}
\begin{figure*}[!ht]
	\begin{center}
		\includegraphics[width=0.85\linewidth,clip=true]{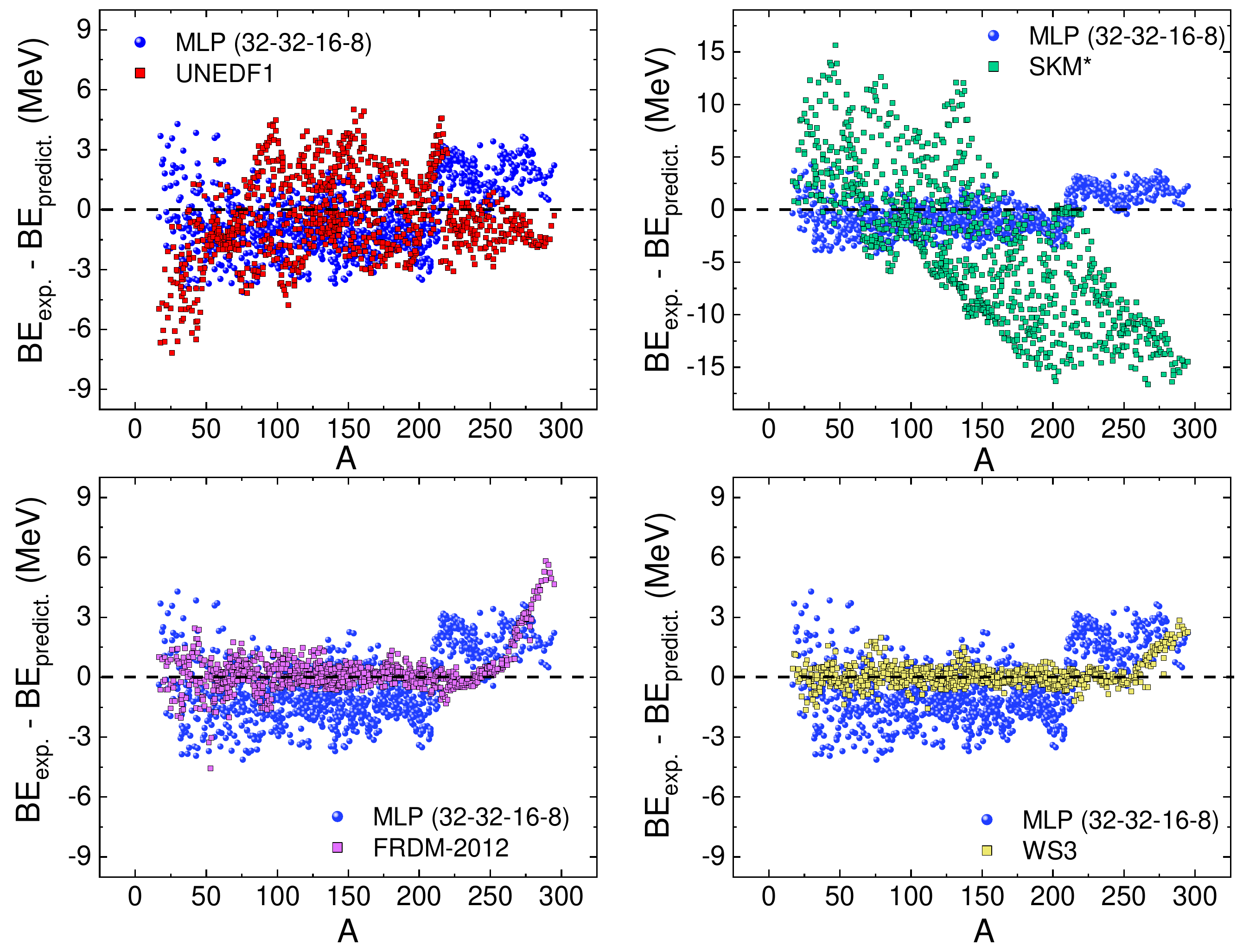}
	\end{center}
	\caption{The same as in Fig. \ref{mass}, but using the four layers MLP architecture.} 
	\label{mass2}
\end{figure*}

It is known that increasing or decreasing the number of hidden layers can also affect the predictive power of the neural networks. Therefore, we also increase the number of the hidden layers to four in the MLP architecture and the same calculations are repeated to see its effect on the results. In Fig. \ref{mass1}, the binding energy differences between the experimental data and MLP predictions with three hidden layers are displayed for 1017 nuclei in the testing set. By increasing the number of the hidden layer by one unit, the predictive power of the MLP model increases, and the $\sigma_{rms}$ values are obtained as 3.72 and 1.84 MeV with two (see Fig.\ref{mass1}(a)) and four inputs (see Fig.\ref{mass1}(b)), respectively. Similar to the MLP with the three hidden layers, the largest deviations in the binding energies are obtained for light and super-heavy nuclei, and inclusion of the additional inputs increases the success of the model in these regions. Using MLP with four hidden layers and two inputs (Z, A), the $\sigma_{rms}$ deviations are obtained as 4.89, 3.06, and 4.82 MeV for light, medium-heavy, and super-heavy nuclei, respectively. Adding the pairing and promiscuity factors to the input channels, the results are improved by about 38.03\%, 48.37\%, and 59.75\% and the $\sigma_{rms}$ deviation values are obtained as 3.03, 1.58, and 1.94 MeV for light, medium-heavy and super-heavy nuclei (see Table \ref{table00}), respectively. 

We should mention that different MLP architectures with different number of hidden layers or hidden units are also tested to make nuclear mass predictions, and obtain the best MPL model. The results of these works are also given in Table \ref{table00}. We found that decreasing the number of the hidden layers also decrease the predictive power of the results. On the other hand, increasing the number of the hidden layers more than four also does not give better results. In this work, the best results are obtained using the three (32-16-8) and four (32-32-16-8) layers MLP architectures with four inputs.
Our results indicate that using the proper number of hidden layers and inputs in the MLP architecture, we can make fast and reliable predictions for nuclear properties. Since the predictions are better using the (Z, A, $\delta$, P) as inputs, we always use the results with four inputs to compare with other mic-mac and microscopic results in the rest of the paper.

In figures \ref{mass} and \ref{mass2}, we also display the binding energy differences between experimental data and modeling results from the MLP architectures with three and four hidden layers and using four (Z, A, $\delta$, P) inputs. Both mic-mac (FRDM-2012 \cite{FRDM2012} and WS3 \cite{WS3}) and microscopic (UNEDF1 \cite{UENDF1,masstables,nature} and SkM* \cite{SKM*,masstables,nature}) results from theoretical database Explorer \cite{masstables} are shown along with the MLP predictions for comparison. It is known that the accuracy of the self-consistent models is not high for the nuclear mass predictions and the root-mean-square deviations are obtained at about several MeV. On the other side, the mic-mac models (FRDM-2012 and WS3) are not self-consistent, and their parameter constants are fitted using the available experimental data, which in turn provide better estimations for the nuclear binding energies as can be seen from Figs. \ref{mass} and \ref{mass2}. The first thing to notice is that the binding energy differences are generally high for light and heavy nuclei in all model predictions. Compared to the UNEDF1, the deviation of the SkM* results from the experimental data is quite high, and increases with the increase in mass number. For mic-mac models (FRDM-2012 and WS3), the highest deviations are obtained for nuclei with A$\geq$250.
Comparing the results of MLP architectures with three and four layers (see figs. \ref{mass} and \ref{mass2}), it is clear that four layers MLP results are better than the three layers MLP.
Besides, both MLP architectures are as successful as the FRDM-2012 and WS3 models in the description of nuclei with A$\geq$250.
Although we only use the proton-mass numbers alongside pairing and promiscuity factors of nuclei as physical inputs in the training of the MLP, the model gives promising results that they are comparable to other microscopic and mic-mac models.

\begin{table}
	\caption{The root-mean square deviations ($\sigma_{rms}$) in units of MeV for the common 956 nuclei using various models. In here, MLP$^{1}$ and MLP$^{2}$ represent three layers (32-16-8) and four layers (32-32-16-8) MLP architectures using (Z, A, $\delta$, P) in the input channel. The nuclear bindings energy results are taken from the nuclear energy density functionals UNEDF1 \cite{UENDF1,masstables,nature} and SkM* \cite{SKM*,masstables,nature}. The results of the FRDM-2012 and WS3 models are taken from Refs. \cite{FRDM2012,WS3}.} 
	\label{table}
	\begin{center}
		%\vspace{6mm}
		\begin{tabular}{cccccccccc}
			\hline\hline 
	& & MLP$^{1}$ & MLP$^{2}$ & UNEDF1 & SKM* & FRDM-2012  & WS3  \\ [1.ex] \cline{3-3} \cline{4-4}\cline{5-5} \cline{6-6} \cline{7-7} \cline{8-8} 
	& $\sigma_{rms}$ & 1.97 & 1.72 & 2.13  & 7.81 & 0.99 & 0.55  				\\ 
			\hline\hline
		\end{tabular}
	\end{center}
	\vspace{-7mm}
\end{table}

In Table \ref{table}, the root-mean-square deviations ($\sigma_{rms}$) are also given for each model to get a better insight into the success of the models. The best results are obtained for FRDM-2012 and WS3 models, and rms deviations are obtained below 1.0 MeV. While the rms deviation is rather high using the SkM* functional and obtained as 7.81 MeV, the UNEDF1 functional makes better predictions and rms deviation is obtained as 2.13 MeV. The root-mean-square deviations are still high using the self-consistent microscopic models in the calculations. Besides, calculations using nuclear energy density functionals are still computationally demanding. It is seen that the MLP model can make reliable predictions that are comparable to the well-known microscopic models, and the $\sigma_{rms}$ values are obtained as 1.97 and 1.72 MeV with three and four layers MLP architectures.

\section{Conclusions}
We implement the multilayer perceptron (MLP) to make ground-state binding energy predictions for atomic nuclei. Two different architectures and inputs are used in the MLP model to study the performance of this neural network in the binding energy predictions. In the first one, we only use the proton and mass numbers of nuclei alongside with the latest experimental data, and no physical input is included. Then, we also added two additional inputs: pairing and promiscuity factors of nuclei to give more physical information to the models. We find that using proper hidden layers and units with relevant information in the input channels, the nuclear binding energy predictions using the MLP improve considerably, especially for light and medium-heavy nuclei. For 1017 nuclei in the testing set, the best root-mean-square deviations are obtained as 2.16 MeV and 1.84 MeV for three and four layers MLP architectures using (Z, A, $\delta$, P) inputs, respectively.

Our findings show that the MLP model can make reasonable predictions for binding energies of atomic nuclei and the results are also comparable to other models. Although the MLP does not include any physics theory behind it and considered as a statistical model, it is seen that the model can make fast and reliable predictions with a proper architecture and relevant inputs. In this respect, the artificial neural networks can be seen as an alternative tool to other mic-mac and microscopic models.

As future work, we plan to extend our calculations by including more physical quantities in the input to better estimate the nuclear properties. Improving the extrapolation abilities of the neural networks for very neutron-rich nuclei is also another challenging task and remains as a future work. Besides, the neural network approaches can be used to train the residues of nuclear properties as it is done in Refs. \cite{PhysRevC.96.044308,NIU201848,PhysRevC.98.034318,PhysRevLett.122.062502}, which in turn can be helpful to understand the missing physics behind the microscopic models.

\bibliography{apssamp}

\end{document}